\documentclass[draft]{agujournal2019}
\usepackage{url} 
\usepackage{amssymb}
\usepackage{lineno}
\usepackage{soul}
\draftfalse

\journalname{Radio Science}

\begin{document}

%
%


\title{GPGPU Acceleration of Incoherent Scatter Radar Plasma Line Analysis}

%
%




\authors{Natalie Hilliard,\affil{1,2}
Juha Vierinen\affil{2},
Philip J. Erickson\affil{2}}


\affiliation{1}{Department of Physics, University of Wisconsin-Madison, Madison, Wisconsin, USA}
\affiliation{2}{Atmospheric Sciences Group, Haystack Observatory, Massachusetts Institute of Technology, Westford, Massachusetts, USA}




\correspondingauthor{Natalie Hilliard}{nvhilliard@gmail.com}



\begin{keypoints}
\item We discuss incoherent scatter radar plasma line echo data analysis and the algorithm that we developed.
\item Our accelerated data pipeline uses graphics processing units to achieve near real-time processing speed.
\end{keypoints}

%
%

%
%


\begin{abstract}
The incoherent scatter radar (ISR) technique is a powerful remote sensing tool for ionosphere and thermosphere dynamics in the near-Earth space enviroment.  
Weak ISR scatter from naturally occuring Langmuir oscillations, or plasma lines, contain high precision information on the altitude-dependent thermal ionospheric electron density.
However, analyzing this frequency-dependent scatter over a large number of radar ranges requires large computational power, especially when the goal is realtime analysis.
General purpose computing on graphics processing units (GPGPU) offers immense computational speedup when compared to traditional central processing unit (CPU) calculations for highly parallelizable tasks, and it is well suited for ISR analysis applications.
This paper extends a single graphics processing unit (GPU) algorithmic solution in a GPGPU framework, and discusses the algorithm developed, including GPU hardware considerations.
Results indicate an order-of-magnitude improvement over CPU analysis and suggest that GPGPU can achieve realtime speed for plasma line applications.
\end{abstract}


%
%

%


%
%
%
%

\section{Introduction}

For more than 50 years, the technique of Incoherent Scatter Radar (ISR) has been developed as a powerful ground-based observational diagnostic that can directly determine the physical properties of Earth's ionosphere.
The ISR technique employs large aperture, high power radars to observe very weak thermal backscatter and obtain full altitude profiles of
such fundamental parameters as ionospheric plasma drift velocities, densities, temperatures, and current densities.  A recent overview of these features is contained in \citeA{Kudeki2012}.

Ionospheric plasma at typical upper atmospheric temperatures and densities contains two primary natural resonances that are used as a remote diagnostic tool.  The first resonance, and the one used most commonly for ISR applications due to its relatively stronger total power compared to background noise, is an inherently narrow band ion-acoustic mode.  The spectral properties of this mode are directly dependent on plasma temperatures, electron and ion densities, and ion compositions. However, application of a robust forward scattering model is needed \cite{Dougherty1960,Farley1961,Dougherty1963}, an absolute radar system calibration needs to be maintained, and ambiguities can exist in the final fitted parameter space.  

A second resonance, and the one focused on computationally in this paper, is the Langmuir mode \cite{Tonks1929}, a high frequency oscillation which is much weaker in power but whose frequency is directly dependent on the square root of electron density with a minor effect from electron temperature in a warm plasma.  The Langmuir mode's plasma density dependence is an attractive target for remote sensing, as it is easier to accurately measure frequency as opposed to power, and the Langmuir mode can provide absolute electron densities as priors for the ion-acoustic fitting task.  For the remainder of this article, we refer to the Langmuir mode as the plasma line.

Until recently, measuring the plasma line in ISR applications was a challenging instrumental task over the full range of its frequency variation, which under typical ionospheric conditions ranges from 1 to 15 MHz above and below the center radar transmitter frequency depending on ambient electron density and temperature.  The plasma line is very narrow, and the combination of this factor with the wide range of plasma line frequencies means that high spectral resolution is required.  The task of computing the required number of spectral bins for full altitude profile determinations has been particularly difficult in general purpose computing hardware when realtime monitoring of plasma line resonances is desired. 

General-purpose computing on graphics processing units (GPGPU) in recent years has developed rapidly as an attractive alternative to traditional CPU processing for computationally expensive tasks.
Utilizing hundreds to thousands of computing cores achieving TFLOPS-single-precision performance, GPUs offer the ideal computational solution to highly-parallelizable problems, often cutting computer time by orders of magnitude over serial CPU code.
These properties have enabled an expansion of GPU usage into many computationally-heavy fields, from biological systems \cite{Dematte2010} to seismic analysis \cite{Michea2010}.
GPU has been readily used to speed up synthetic aperture radar data processing \cite{Rubin2010, Ning2011}, but it seems rather underutilized, or at least underdocumented, in ISR applications, and is only briefly discussed in the current literature by \citeA{Fallen2011}.

NVIDIA GPU's are the common choice for scientific GPGPU implementation, largely aided by the established and mature development community.
Part of this success is the ease of use, provided by NVIDIA's Compute Unified Device Architecture (CUDA), it's API, and various accompanying libraries \cite{CUDA2015}.
This programming model utilizes high level programming languages, such as C, with minimal extensions to provide access to GPU-specific hardware and computation \cite{PrgGuide2015}.

In this paper, we discuss current ISR plasma line analysis techniques and their implementation on CPU, including the speed difficulties faced.
We then discuss GPGPU and hardware considerations necessary for a robust GPU algorithm, and discuss the algorithm developed.
The performance of this algorithm is discussed and compared to previous CPU results.
Demonstrations here suggest that sufficiently powerful single GPU or cheaper, multiple GPU solutions could achieve realtime goals.


\section{ISR Data and Processing}

\subsection{Plasma Lines}
The ion acoustic line is determined to first order by the Doppler shifts associated with the thermal velocity distribution of ions. 
Due to the large masses involved, this feature is relatively narrow in frequency ($< 50$ kHz). 
The plasma line on the other hand is a narrow line, but it is offset from the ion line by the plasma resonance frequency $\pm f_r = \omega_r/2\pi$, which is determined to first order in \citeA{Yngvesson1968}: 
\begin{equation}
f_r^2 = f_p^2 + \frac{3k^2}{4\pi^2}\frac{k_bT_e}{m_e} + f_c^2\sin^2(\alpha)
\end{equation}
Here, $f_p$ is the plasma frequency, $k_b$ is Boltzmann's constant, $k$ is the wavenumber, $T_e$ is the electron temperature, $m_e$ is the electron mass, $f_c$ is electron gyro frequency, and $\alpha$ is the aspect angle between the wave vector and magnetic field.
This can range anywhere between 1 and 15 MHz for typical ionospheric conditions. Because the ionospheric electron density from minima to maxima ranges through these values, in order to measure a full incoherent scatter spectrum, one need to record radar echoes with a bandwidth of approximately 30 MHz. 

The analysis of ion lines and plasma lines can be performed using a method called lag-profile inversion \cite{Virtanen2008}. 
This method especially useful for ion-line work. 
However, for high resolution plasma line measurements, where we need to measure the spectrum with a range resolution of 150 meters and frequency resolution of ~1 kHz, a faster numerical method is needed. 
In this paper we will focus on such a faster method, which makes use of the fact that a matched filter is nearly equivalent to a maximum likelihood estimate for a certain set of codes. 

\subsection{Plasma Line Signal Processing}

Our primary concern here is resolving the power spectrum from the raw radar echo. To do this, we start with the measurement equation giving the measured echo signal at time t:
\begin{equation}
m_t = \displaystyle\sum_{r} \displaystyle\sum_{\omega} \epsilon_{t-r}
e^{i\omega t}\sigma_{r,\omega} + \xi_{t}
\end{equation}
Here, $r$ is the range ($ct$), $\epsilon_{t-r}$ is the transmit pulse, $\omega$ is its frequency, $\sigma$ is the received scatter, and $\xi$ is the environmental noise.
Due to the incoherence of ISR scatter, the return scatter is a random complex number from a Gaussian distribution centered at 0 with deviation of power spectral density $S_{r,\omega}$.
While the echo $\sigma_{r,\omega}$ does not change for a singular pulse traveling through the target, it does change from pulse to pulse; however, $S_{r,\omega}$ doesn't.
\begin{equation}
\sigma_{r,\omega} = N_{\mathbb{C}}(0,S_{r,\omega})
\end{equation}
We can rewrite this in linear algebraic form with computation now smuggled into the A transform:
\begin{equation}
m = A x + \xi
\end{equation}
The maximum likelihood estimate $x_{ML}$ cannot be solved because there are more unknowns than measurements -- the problem is undetermined, and hence we cannot use the standard formula:
\begin{equation}
x_{ML} = (A^HA)^{-1}A^Hm
\end{equation}
However, with an additional assumption that the transmit pulses have such a waveform that the following property is valid:
\begin{equation}
A^HA \propto \alpha I
\end{equation}
i.e., the Fischer information matrix is close to an identity matrix up to a scalar constant.
Using this property, we can approximate the maximum likelihood estimate as the correlation estimate $x_{CE}$:
\begin{equation}
x_{ML} \approx A^Hm=x_{CE}
\end{equation}
And this, squared, gives us the power spectrum that we desire:
\begin{equation}
P = \langle |x_{CE}|^2 \rangle
\end{equation}
where vector $P$ contains $S_{r,\omega}$ as elements.

\begin{figure*}[ht]
\centering
\noindent\includegraphics[width=32pc]{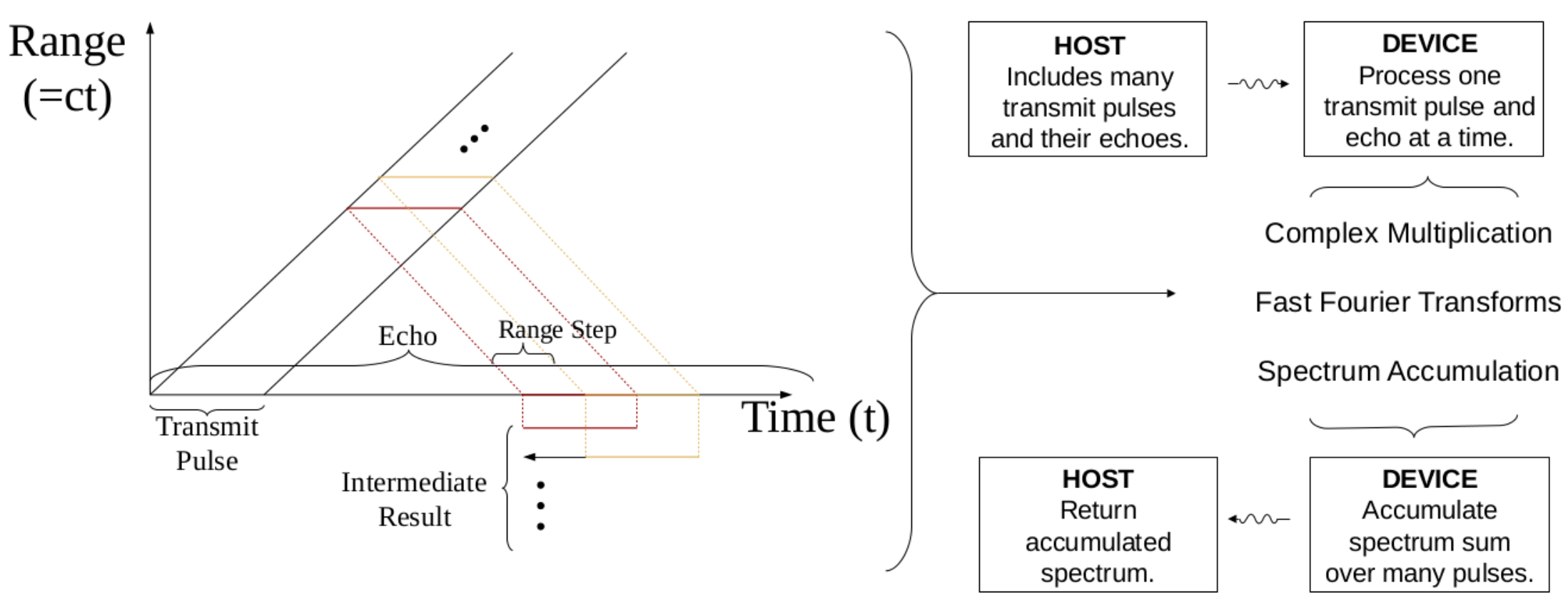}
\caption{Visualization of the GPU processing algorithm developed and the space-time diagram of the radar pulses.}
\label{spacetime}
\end{figure*}

As a result, we can process the measurement transmit pulses and echoes independently, allowing GPU parallelization.
The steps required for processing (packed away in $A^H$ above) start with a complex conjugation of the transmit pulse, a complex multiplication of this into one range gate of the return echo, and a Fourier transform of that result.
We square this to arrive at the power for the single range gate used.
Repeating this by iterating through the number of range gates we desire and the amount to step each of them, we process the entire echo and return a power spectrum, seen in Figure \ref{spacetime}.

\section{GPU Acceleration}

\subsection{Hardware}

In GPGPU, data is transferred from the host (CPU) to device (GPU) memory, calculations are run using the GPU, and the result is then transferred back to the host.
In order to achieve fastest computation time, we need to minimize the relatively slow data transfer between host and device (8 GB/s for PCIe 1.x x16) and maximize utilization and throughput on the GPU.

Each NVIDIA GPU consists of several streaming multiprocessors (SMs), each containing several CUDA cores, specified by the device's compute capability, a number describing the architecture base and other device characteristics.
When instructions, or kernels, are passed to the GPU, they are passed in units called thread blocks, or blocks, in one, two, or three dimensions.
These blocks can be ordered as a grid of one or two dimensions to sufficiently break up large data.
Blocks are queued and executed in succession on the SMs so that GPUs with larger numbers of SMs will compute the same tasks, albeit at a faster rate \cite{PrgGuide2015}. 

Parameters specified in the kernel call determine the amount of parallelism, declaring N executions of by N threads in parallel.
These threads comprise each of the blocks to be processed by each SM.
Within each SM (and block), CUDA cores process the threads in parallel, grouped as warps.
Each warp (32 threads each for current GPUs) is executed in the order dictated by the warp scheduler.
GPU SMs break the blocks into warps, ideally having the same execution path within each warp, and it is advantageous for the block size to be a multiple of the warp size for maximum utilization.
Threads within a warp also coordinate coalesced access of global memory, so it is also advantageous for each batch of data be continuous \cite{PrgGuide2015}.

\subsection{Accelerated Algorithm}

In order to implement the desired CPU algorithm efficiently, we find it necessary to break the GPU implementation into separate kernels.
We choose this to take advantage of the optimized cuFFT library for the fast Fourier transforms (FFTs) used \cite{cuFFT2015}.
Because of this, we describe each kernel separately and will summarize their use together as a complete algorithm below.

\subsubsection{Complex Multiplication}

This custom kernel takes the conjugate of the transmit signal (TX) and complex multiplies that with the corresponding element in the return echo, writing the result to a row of batch (intermediate output within GPU memory) for each range gate.
Each thread is writing to a linearized index of batch according to its given block and thread location.
Using this linearized index with the signal length (row length), the thread calculates the 2-D coordinate position of its write in batch.
The first dimension coordinate specifies which range gate to start from, and the second specifies the element along the TX.
With these two coordinates, the thread determines the element needed from echo that corresponds to each TX and batch element.
This method preserves memory continuity required for efficient global memory access.

\subsubsection{Fast Fourier Transform}

Next, we need an efficient method to calculate the FFTs desired.
To this end, cuFFT fulfilled these goals, offering a highly optimized, simple interface to efficiently calculate the transforms on the GPU.
This method allows us to quickly implement the GPU FFTs with high parallelism without writing a custom kernel.
We choose to use a 1-D FFT operating on each row of the batch output from the previous kernel, returning the solutions in-place.
In order to improve computation speed, we set the rows to have lengths as a power of two (16384).

\subsubsection{Spectrum Accumulation}

Our third calculation uses another custom kernel to take the square of the complex values returned in batch and adding these to a new spectrum floating point array.
The same linearized mapping used in the complex multiplication is used again here, so each thread will add the squared complex value to the corresponding spectrum element.
This implementation allows us to loop outside of the kernel to accumulate a spectrum for a number of radar pulses.

\subsubsection{Combined Algorithm}

To reduce memory transfers between host and device, we move a single full TX and echo over to the device, then execute each kernel in succession.
Intermediate results are kept in the device memory, and only the final spectrum is transferred back to the host memory.
Before the kernels can be called, we set up the block and grid dimensions to canvas all the elements in the intermediate batch array.
Respecting the warp specifications of the GPU, we set blocks as 16x16.
These dimensions are efficient and portable, transferring well to other NVIDIA GPUs while maintaining high computation speed.
Grid dimensions are determined with division of these block dimensions into the data size.
We assume the data size to be evenly divisible by the block dimensions; for cases where this does not hold, we recommend padding the data.
Otherwise, it will be necessary to add a catch in each custom kernel (slowing computation) ensuring the indexing does not overrun the data in the final block(s).

The host data, however, contains numerous pulses and echoes.
To process all of the required pulses, we copy and process one row at a time, accumulating the results into the device spectrum.
Upon completion, we transfer the spectrum result back to the host memory.


\section{Results}

To test the computation speed of the algorithms, we use 100 simulated signals and echoes of 16,384 and 250,000 elements respectively, created with the SciPy toolkit \cite{scipy2015}.
We use 4096 range gates with a step size of 25, starting at the first element.

For the GPU tests, we wrap the CUDA C function in Python, simulate the data, and feed the simulated data into the wrapped function.
Within the C function, timing includes only the row transfers from host memory to device memory and the kernel executions on each of the 100 rows.

For the CPU test, we immediately run NumPy operations on the simulated data, applying the same operations as those in the CUDA code.
The algorithm takes advantage of NumPy vector operations and FFTs to provide a reasonably competitive comparison to GPU tests. 
Despite having no required memory transfers, this host-calculated code shows the order-of-magnitude speed upgrade of the GPU algorithm, evidenced by Table \ref{results}.
\begin{table*}[ht]
{\small
\hfill{}
\begin{tabular}{l c r r r}
\hline
\textbf{Hardware} & \textbf{Bandwidth(GB/s)} & \textbf{Time(s)} & \textbf{Speed Ratio} & \textbf{Speedup}\\
\hline
  Intel i5-4660 &  N/A & 120 & 0.008  &   1.00x \\
  TESLA C2050   &  8.0 &   4.423 & 0.226  &  27.13x \\
  GTX 970       &  8.0 &   2.530 & 0.395  &  47.43x \\
  GTX TITAN     &  8.0 &   1.742 & 0.574  &  68.89x \\
  GTX 780 Ti & 8.0 & 1.488 & 0.672 & 80.65x \\
\hline
\end{tabular}}
\hfill{}
\caption{Simulated data processing time, host to device bandwidth (two-way), realtime speed ratio, and comparison to CPU compute time of assorted processors.}
\label{results}
\end{table*}

We test the GPU algorithm on different systems, with relevant system information also included in Table \ref{results}.
No code adjustments were required for these separate systems, highlighting the portability of our algorithm.
As the radar echoes at Millstone Hill are collected every 10ms, we compare the plasma line processing to this acquisition rate to illustrate the computation speed in an intuitive manner.


\section{Conclusion}

We have discussed the successful application of a highly-parallelized GPU algorithm to ISR plasma line analysis.
Our GPGPU implementation offers order-of-magnitude speedup over traditional CPU computation, achieving near realtime computation of plasma line data on an appropriately powerful single GPU.
Thus, the GPU algorithm presented here presents an attractive and economical choice for ISR plasma line processing, with good portability and flexibility for general use.

Our algorithm is not the upper limit for plasma line computation.
Additional optimization is possible, and GPGPU technology is advancing rapidly; it is likely that together, these two factors can bring plasma line analysis to realtime speed with a cheap, single GPU solution.
Possible optimization avenues include but are not limited to: (1) decreasing block sizes and using shared memory with strided thread computation (while respecting memory coalescence); (2) asynchronous pinned memory copying from host to device, occurring during kernel executions; and (3) dynamically reducing echo size to only transfer necessary elements without significant overhead.

\section*{Open Research Section}
Our program is available at https://doi.org/10.5281/zenodo.10215279

\acknowledgments
Special thanks to MIT Haystack Observatory for project support, including technical expertise, facilities, and hardware. 
Extra thanks to project mentors Juha Vierinen and Phil Erickson for their direction and guidance during project development.
This was made possible by the NSF Research Experiences for Undergraduates grant AST-1156504 to MIT Haystack Observatory.

%
%

\bibliography{agu}

\begin{thebibliography}{}

\bibitem [\protect \citeauthoryear {%
Dematt\'{e}%
\ \BBA {} Prandi%
}{%
Dematt\'{e}%
\ \BBA {} Prandi%
}{%
{\protect \APACyear {2010}}%
}]{%
Dematte2010}
\APACinsertmetastar {%
Dematte2010}%
\begin{APACrefauthors}%
Dematt\'{e}, L.%
\BCBT {}\ \BBA {} Prandi, D.%
\end{APACrefauthors}%
\unskip\
\newblock
\APACrefYearMonthDay{2010}{}{}.
\newblock
{\BBOQ}\APACrefatitle {{GPU computing for systems biology}} {{GPU computing for systems biology}}.{\BBCQ}
\newblock
\APACjournalVolNumPages{Briefings in Bioinformatics}{11}{3}{323--333}.
\newblock
\begin{APACrefDOI} \doi{10.1093/bib/bbq006} \end{APACrefDOI}
\PrintBackRefs{\CurrentBib}

\bibitem [\protect \citeauthoryear {%
Dougherty%
\ \BBA {} Farley%
}{%
Dougherty%
\ \BBA {} Farley%
}{%
{\protect \APACyear {1960}}%
}]{%
Dougherty1960}
\APACinsertmetastar {%
Dougherty1960}%
\begin{APACrefauthors}%
Dougherty, J\BPBI P.%
\BCBT {}\ \BBA {} Farley, D\BPBI T.%
\end{APACrefauthors}%
\unskip\
\newblock
\APACrefYearMonthDay{1960}{{\APACmonth{11}}}{}.
\newblock
{\BBOQ}\APACrefatitle {{A Theory of Incoherent Scattering of Radio Waves by a Plasma}} {{A Theory of Incoherent Scattering of Radio Waves by a Plasma}}.{\BBCQ}
\newblock
\APACjournalVolNumPages{Proceedings of the Royal Society of London. Series A}{259}{}{79}.
\newblock
\begin{APACrefURL} \url{http://adsabs.harvard.edu/cgi-bin/nph-data\_query?bibcode=1960RSPSA.259...79D\&link\_type=ABSTRACT papers3://publication/uuid/9EE43795-E0E9-47B5-BD3B-C1EBAF51FBE0} \end{APACrefURL}
\PrintBackRefs{\CurrentBib}

\bibitem [\protect \citeauthoryear {%
Dougherty%
\ \BBA {} Farley%
}{%
Dougherty%
\ \BBA {} Farley%
}{%
{\protect \APACyear {1963}}%
}]{%
Dougherty1963}
\APACinsertmetastar {%
Dougherty1963}%
\begin{APACrefauthors}%
Dougherty, J\BPBI P.%
\BCBT {}\ \BBA {} Farley, D\BPBI T.%
\end{APACrefauthors}%
\unskip\
\newblock
\APACrefYearMonthDay{1963}{{\APACmonth{10}}}{}.
\newblock
{\BBOQ}\APACrefatitle {{A Theory of Incoherent Scattering of Radio Waves by a Plasma, 3 Scattering in a Partly Ionized Gas}} {{A Theory of Incoherent Scattering of Radio Waves by a Plasma, 3 Scattering in a Partly Ionized Gas}}.{\BBCQ}
\newblock
\APACjournalVolNumPages{Journal of Geophysical Research}{68}{}{5473}.
\newblock
\begin{APACrefURL} \url{http://adsabs.harvard.edu/cgi-bin/nph-data\_query?bibcode=1963JGR....68.5473D\&link\_type=CITATIONS papers3://publication/uuid/004D5C25-61E1-42C0-8F54-E2CC28688D98} \end{APACrefURL}
\PrintBackRefs{\CurrentBib}

\bibitem [\protect \citeauthoryear {%
Fallen%
, Bellamy%
, Newby%
\BCBL {}\ \BBA {} Watkins%
}{%
Fallen%
\ \protect \BOthers {.}}{%
{\protect \APACyear {2011}}%
}]{%
Fallen2011}
\APACinsertmetastar {%
Fallen2011}%
\begin{APACrefauthors}%
Fallen, C.%
, Bellamy, B.%
, Newby, G.%
\BCBL {}\ \BBA {} Watkins, B.%
\end{APACrefauthors}%
\unskip\
\newblock
\APACrefYearMonthDay{2011}{}{}.
\newblock
{\BBOQ}\APACrefatitle {GPU Performance Comparison for Accelerated Radar Data Processing} {Gpu performance comparison for accelerated radar data processing}.{\BBCQ}
\newblock
\BIn{} \APACrefbtitle {2011 Symposium on Application Accelerators in High-Performance Computing} {2011 symposium on application accelerators in high-performance computing}\ (\BPG~84-92).
\newblock
\begin{APACrefDOI} \doi{10.1109/SAAHPC.2011.14} \end{APACrefDOI}
\PrintBackRefs{\CurrentBib}

\bibitem [\protect \citeauthoryear {%
Farley%
, Dougherty%
\BCBL {}\ \BBA {} Barron%
}{%
Farley%
\ \protect \BOthers {.}}{%
{\protect \APACyear {1961}}%
}]{%
Farley1961}
\APACinsertmetastar {%
Farley1961}%
\begin{APACrefauthors}%
Farley, D\BPBI T.%
, Dougherty, J\BPBI P.%
\BCBL {}\ \BBA {} Barron, D\BPBI W.%
\end{APACrefauthors}%
\unskip\
\newblock
\APACrefYearMonthDay{1961}{{\APACmonth{09}}}{}.
\newblock
{\BBOQ}\APACrefatitle {{A Theory of Incoherent Scattering of Radio Waves by a Plasma II. Scattering in a Magnetic Field}} {{A Theory of Incoherent Scattering of Radio Waves by a Plasma II. Scattering in a Magnetic Field}}.{\BBCQ}
\newblock
\APACjournalVolNumPages{Proceedings of the Royal Society of London. Series A}{263}{}{238}.
\newblock
\begin{APACrefURL} \url{http://adsabs.harvard.edu/cgi-bin/nph-data\_query?bibcode=1961RSPSA.263..238F\&link\_type=ABSTRACT papers3://publication/uuid/FB09372F-34C3-49AC-80A9-D41764F6E6A1} \end{APACrefURL}
\PrintBackRefs{\CurrentBib}

\bibitem [\protect \citeauthoryear {%
Jones%
, Oliphant%
, Peterson%
\BCBL {}\ \protect \BOthers {.}}{%
Jones%
\ \protect \BOthers {.}}{%
{\protect \APACyear {2001--}}%
}]{%
scipy2015}
\APACinsertmetastar {%
scipy2015}%
\begin{APACrefauthors}%
Jones, E.%
, Oliphant, T.%
, Peterson, P.%
\BCBL {}\ \BOthersPeriod {.}\end{APACrefauthors}%
\unskip\
\newblock
\APACrefYearMonthDay{2001--}{}{}.
\newblock
\APACrefbtitle {{SciPy}: Open source scientific tools for {Python}.} {{SciPy}: Open source scientific tools for {Python}.}
\newblock
\begin{APACrefURL} \url{http://www.scipy.org/} \end{APACrefURL}
\newblock
\APACrefnote{[Online; accessed 2015-07-27]}
\PrintBackRefs{\CurrentBib}

\bibitem [\protect \citeauthoryear {%
Kudeki%
\ \BBA {} Milla%
}{%
Kudeki%
\ \BBA {} Milla%
}{%
{\protect \APACyear {2012}}%
}]{%
Kudeki2012}
\APACinsertmetastar {%
Kudeki2012}%
\begin{APACrefauthors}%
Kudeki, E.%
\BCBT {}\ \BBA {} Milla, M.%
\end{APACrefauthors}%
\unskip\
\newblock
\APACrefYearMonthDay{2012}{}{}.
\newblock
{\BBOQ}\APACrefatitle {{16 Incoherent Scatter Radar — Spectral Signal Model and Ionospheric Applications}} {{16 Incoherent Scatter Radar — Spectral Signal Model and Ionospheric Applications}}.{\BBCQ}
\newblock
\APACjournalVolNumPages{Doppler Radar Observations - Weather Radar, Wind Profiler, Ionospheric Radar, and Other Advanced Applications}{}{}{377--406}.
\newblock
\begin{APACrefDOI} \doi{10.5772/2036} \end{APACrefDOI}
\PrintBackRefs{\CurrentBib}

\bibitem [\protect \citeauthoryear {%
Mich\'{e}a%
\ \BBA {} Komatitsch%
}{%
Mich\'{e}a%
\ \BBA {} Komatitsch%
}{%
{\protect \APACyear {2010}}%
}]{%
Michea2010}
\APACinsertmetastar {%
Michea2010}%
\begin{APACrefauthors}%
Mich\'{e}a, D.%
\BCBT {}\ \BBA {} Komatitsch, D.%
\end{APACrefauthors}%
\unskip\
\newblock
\APACrefYearMonthDay{2010}{}{}.
\newblock
{\BBOQ}\APACrefatitle {{Accelerating a three-dimensional finite-difference wave propagation code using GPU graphics cards}} {{Accelerating a three-dimensional finite-difference wave propagation code using GPU graphics cards}}.{\BBCQ}
\newblock
\APACjournalVolNumPages{Geophysical Journal International}{182}{1}{389--402}.
\newblock
\begin{APACrefDOI} \doi{10.1111/j.1365-246X.2010.04616.x} \end{APACrefDOI}
\PrintBackRefs{\CurrentBib}

\bibitem [\protect \citeauthoryear {%
Ning%
, Yeh%
, Zhou%
, Gao%
\BCBL {}\ \BBA {} Yang%
}{%
Ning%
\ \protect \BOthers {.}}{%
{\protect \APACyear {2011}}%
}]{%
Ning2011}
\APACinsertmetastar {%
Ning2011}%
\begin{APACrefauthors}%
Ning, X.%
, Yeh, C.%
, Zhou, B.%
, Gao, W.%
\BCBL {}\ \BBA {} Yang, J.%
\end{APACrefauthors}%
\unskip\
\newblock
\APACrefYearMonthDay{2011}{}{}.
\newblock
{\BBOQ}\APACrefatitle {{Multiple-GPU accelerated range-Doppler algorithm for synthetic aperture radar imaging}} {{Multiple-GPU accelerated range-Doppler algorithm for synthetic aperture radar imaging}}.{\BBCQ}
\newblock
\APACjournalVolNumPages{IEEE National Radar Conference - Proceedings}{}{}{698--701}.
\newblock
\begin{APACrefDOI} \doi{10.1109/RADAR.2011.5960627} \end{APACrefDOI}
\PrintBackRefs{\CurrentBib}

\bibitem [\protect \citeauthoryear {%
{NVIDIA Corporation}%
}{%
{NVIDIA Corporation}%
}{%
{\protect \APACyear {2015}}%
{\protect \APACexlab {{\protect \BCnt {1}}}}}]{%
PrgGuide2015}
\APACinsertmetastar {%
PrgGuide2015}%
\begin{APACrefauthors}%
{NVIDIA Corporation}.%
\end{APACrefauthors}%
\unskip\
\newblock
\APACrefYearMonthDay{2015{\protect \BCnt {1}}}{}{}.
\newblock
\APACrefbtitle {{CUDA Programming Guide Version 7.0}.} {{CUDA Programming Guide Version 7.0}.}
\newblock
\begin{APACrefURL} \url{http://docs.nvidia.com/cuda/cuda-c-programming-guide/} \end{APACrefURL}
\PrintBackRefs{\CurrentBib}

\bibitem [\protect \citeauthoryear {%
{NVIDIA Corporation}%
}{%
{NVIDIA Corporation}%
}{%
{\protect \APACyear {2015}}%
{\protect \APACexlab {{\protect \BCnt {2}}}}}]{%
CUDA2015}
\APACinsertmetastar {%
CUDA2015}%
\begin{APACrefauthors}%
{NVIDIA Corporation}.%
\end{APACrefauthors}%
\unskip\
\newblock
\APACrefYearMonthDay{2015{\protect \BCnt {2}}}{}{}.
\newblock
\APACrefbtitle {{CUDA Technology Version 7.0}.} {{CUDA Technology Version 7.0}.}
\newblock
\begin{APACrefURL} \url{http://www.nvidia.com/object/cuda\_home\_new.html} \end{APACrefURL}
\PrintBackRefs{\CurrentBib}

\bibitem [\protect \citeauthoryear {%
{NVIDIA Corporation}%
}{%
{NVIDIA Corporation}%
}{%
{\protect \APACyear {2015}}%
{\protect \APACexlab {{\protect \BCnt {3}}}}}]{%
cuFFT2015}
\APACinsertmetastar {%
cuFFT2015}%
\begin{APACrefauthors}%
{NVIDIA Corporation}.%
\end{APACrefauthors}%
\unskip\
\newblock
\APACrefYearMonthDay{2015{\protect \BCnt {3}}}{}{}.
\newblock
\APACrefbtitle {{NVIDIA CUDA cuFFT Library 7.0}.} {{NVIDIA CUDA cuFFT Library 7.0}.}
\newblock
\begin{APACrefURL} \url{http://docs.nvidia.com/cuda/cufft/index.html\#axzz3gj9AOXQX} \end{APACrefURL}
\PrintBackRefs{\CurrentBib}

\bibitem [\protect \citeauthoryear {%
Rubin%
, Sager%
, Ph%
\BCBL {}\ \BBA {} Berger%
}{%
Rubin%
\ \protect \BOthers {.}}{%
{\protect \APACyear {{\protect \bibnodate {}}}}%
}]{%
Rubin2010}
\APACinsertmetastar {%
Rubin2010}%
\begin{APACrefauthors}%
Rubin, G.%
, Sager, E\BPBI V.%
, Ph, D.%
\BCBL {}\ \BBA {} Berger, D\BPBI H.%
\end{APACrefauthors}%
\unskip\
\newblock
\APACrefYearMonthDay{{\protect \bibnodate {}}}{}{}.
\newblock
{\BBOQ}\APACrefatitle {{GPU Acceleration of SAR / ISAR Imaging Algorithms}} {{GPU Acceleration of SAR / ISAR Imaging Algorithms}}.{\BBCQ}
\newblock
\APACjournalVolNumPages{}{}{}{2--6}.
\PrintBackRefs{\CurrentBib}

\bibitem [\protect \citeauthoryear {%
Tonks%
\ \BBA {} Langmuir%
}{%
Tonks%
\ \BBA {} Langmuir%
}{%
{\protect \APACyear {1929}}%
}]{%
Tonks1929}
\APACinsertmetastar {%
Tonks1929}%
\begin{APACrefauthors}%
Tonks, L.%
\BCBT {}\ \BBA {} Langmuir, I.%
\end{APACrefauthors}%
\unskip\
\newblock
\APACrefYearMonthDay{1929}{}{}.
\newblock
{\BBOQ}\APACrefatitle {{Oscillations in ionized gases}} {{Oscillations in ionized gases}}.{\BBCQ}
\newblock
\APACjournalVolNumPages{Physical Review}{33}{2}{195--210}.
\newblock
\begin{APACrefDOI} \doi{10.1103/PhysRev.33.195} \end{APACrefDOI}
\PrintBackRefs{\CurrentBib}

\bibitem [\protect \citeauthoryear {%
Virtanen%
, Lehtinen%
, Nygr\'{e}n%
, Orisp\"{a}\"{a}%
\BCBL {}\ \BBA {} Vierinen%
}{%
Virtanen%
\ \protect \BOthers {.}}{%
{\protect \APACyear {2008}}%
}]{%
Virtanen2008}
\APACinsertmetastar {%
Virtanen2008}%
\begin{APACrefauthors}%
Virtanen, I\BPBI I.%
, Lehtinen, M\BPBI S.%
, Nygr\'{e}n, T.%
, Orisp\"{a}\"{a}, M.%
\BCBL {}\ \BBA {} Vierinen, J.%
\end{APACrefauthors}%
\unskip\
\newblock
\APACrefYearMonthDay{2008}{}{}.
\newblock
{\BBOQ}\APACrefatitle {{Lag profile inversion method for EISCAT data analysis}} {{Lag profile inversion method for EISCAT data analysis}}.{\BBCQ}
\newblock
\APACjournalVolNumPages{Annales Geophysicae}{26}{3}{571--581}.
\newblock
\begin{APACrefDOI} \doi{10.5194/angeo-26-571-2008} \end{APACrefDOI}
\PrintBackRefs{\CurrentBib}

\bibitem [\protect \citeauthoryear {%
Yngvesson%
\ \BBA {} Perkins%
}{%
Yngvesson%
\ \BBA {} Perkins%
}{%
{\protect \APACyear {1968}}%
}]{%
Yngvesson1968}
\APACinsertmetastar {%
Yngvesson1968}%
\begin{APACrefauthors}%
Yngvesson, K\BPBI O.%
\BCBT {}\ \BBA {} Perkins, F\BPBI W.%
\end{APACrefauthors}%
\unskip\
\newblock
\APACrefYearMonthDay{1968}{}{}.
\newblock
\APACrefbtitle {{Radar Thomson scatter studies of photoelectrons in the ionosphere and Landau damping}} {{Radar Thomson scatter studies of photoelectrons in the ionosphere and Landau damping}}\ (\BVOL~73)\ (\BNUM~1).
\newblock
\begin{APACrefDOI} \doi{10.1029/JA073i001p00097} \end{APACrefDOI}
\PrintBackRefs{\CurrentBib}

\end{thebibliography}

%
%
%
%
%

\end{document}